\documentclass{moriond}

\def\be{\begin{equation}}
\def\ee{\end{equation}}
\def\bea{\begin{eqnarray}}
\def\eea{\end{eqnarray}}

\usepackage{lineno}  

\newcommand{\Photo}{}

\begin{document}
\vspace*{4cm}
\title{B AND C HADRON SPECTROSCOPY AT LHCB}

\author{ I. POLYAKOV (FOR THE LHCB COLLABORATION) } 

\address{Institute for Theoretical and Experimental Physics (ITEP), Moscow, Russia}

\maketitle\abstracts{
Recent results in the field of $b$ and $c$ hadron spectroscopy at the LHCb experiment are presented. The analyses use the data collected with the LHCb detector in proton-proton collisions corresponding to an integrated luminosity up to 3.0 fb$^{-1}$.}

\section{Introduction}

The LHCb detector~\cite{LHCb_detector} is a single-arm forward spectrometer at the LHC, designed for the study of heavy flavour physics.
The unique geometry of the LHCb experiment accepts 40\% of all $b\overline{b}$ pairs produced in proton-proton collisions. 
Together with efficient trigger, excellent momentum and decay time resolution and particle identification,
this allows to perform studies of hadrons containing $b$ or $c$ quarks with high precision.

\section{Search for excited $D_J$ mesons}

Charm meson spectroscopy allows to test predictions of the quark model. 
Only few of the predicted states are well established. Many states have not been observed yet or need to be confirmed. 
A search for $D_{J}$ mesons in $D^+\pi^-$, $D^0\pi^+$ and $D^{*+}\pi^-$ final states is reported~\cite{Dj}.

Combined with the fit to the invariant mass of $D_J$ candidates, angular analysis in the ${D_{J}\rightarrow D^{*+}\pi^-}$ decay mode allows to distinguish between natural and unnatural parity~\footnote{The states having $P=(-1)^{J}$ and therefore spin-parity ${J^{P}=0^{+},1^{-},2^{+},...}$ are called natural and are labeled as $D^{*}$, while the states with another parity ($J^{P}=0^{-},1^{+},2^{-},...$) are called unnatural.} states. The angle between the $\pi^+$ from the $D^{*+}$ decay and the $\pi^-$, determined in the $D_{J}$ rest frame, is used to separate between different spin-parity components.

In addition to the well-established $D_1(2420)^0$ and $D^*(2460)^0$ states, seven high-mass resonances are observed.
In the $D^{*+}\pi^-$ final state two natural parity states, $D^*_J(2650)^0$ and $D^*_J(2760)^0$, and three unnatural parity states, $D_J(2580)^0$, $D_J(2740)^0$ and $D_J(3000)^0$, are seen and a measurement of their masses and widths is performed. The $D^*_J(2760)$ state is also observed in the $D^+\pi^-$ and $D^0\pi^+$ final states, while for the region of $D^*_J(2650)$ no conclusive statement can be made due to cross-feed of partially reconstructed $D_J$ decays. In the $D^+\pi^-$ and $D^0\pi^+$ mass spectra $D^*_J(3000)^0$ and $D^*_J(3000)^+$ structures are also observed. These structures could be superpositions of 1F states expected by the quark model. 

\section{Search for $\Xi_{cc}^{+}$ baryon}

All of the baryon ground states with charm quantum number $C=0,1$ have been discovered. Three states with $C=2$ are expected: a $\Xi_{cc}$ isodoublet and an $\Omega_{cc}$ isosinglet. 
The SELEX experiment has claimed the observation of the $\Xi_{cc}^+$ state in $\Lambda_c^+ K^-\pi^+$ and $p D^+ K^-$ decay modes~\cite{SELEX}. 
The mass was measured to be $3519 \pm 2$~MeV/c$^2$ and for the lifetime an upper limit of $33$~fs was determined, while the theoretical calculations predict the lifetime to be between 100~fs and 250~fs. Later searches for the $\Xi_{cc}^+$ by FOCUS, BaBar and Belle experiments have not confirmed this observation. 

The search for the $\Xi_{cc}^+\rightarrow \Lambda_c^+ K^-\pi^+$ is reported~\cite{Xicc}. The $\Xi_{cc}^+$ is searched in a range of mass between 3300~MeV/c$^2$ and 3800~MeV/c$^2$ and lifetime between 100~fs and 400~fs, which correspond to values expected from the theory. No signal has been found, therefore upper limits on 
\begin{equation}
R \equiv \frac{\sigma(\Xi_{cc}^+)\times {\cal B}(\Xi_{cc}^+\rightarrow \Lambda_c^+ K^-\pi^+)}{\sigma(\Lambda_c^+)} 
\end{equation} 
\noindent are obtained for a range of mass and lifetime hypotheses. The upper limit strongly depends on the assumed lifetime varying from $R<1.5\times 10^{-2}$ for 100~fs to $R<3.9\times 10^{-4}$ for 400~fs. However, these upper limits do not exclude the SELEX result due to the possibility of very short $\Xi_{cc}^+$ lifetime ($\ll$100~fs) or differences in production environment. 

\section{Evidence for the decay $X(3872)\rightarrow\psi(2S)\gamma$}

The $X(3872)$ state was discovered in 2003 by the Belle collaboration~\cite{belleX} 
and later was observed by many other experiments. Despite of large amount of experimental information, its nature is still unclear with many possible interpretations such as conventional charmonium $\chi_{c1}(2P)$ state, $D\overline{D}^*$ molecule, tetraquark, hybrid meson, glueball or their mixtures. 
A measurement of the ratio $R_{\psi\gamma}={\cal B}(X(3872)\rightarrow\psi(2S)\gamma)/{\cal B}(X(3872)\rightarrow J/\psi\gamma)$ could help to distinguish between these possibilities, as the theory predictions for this quantity vary widely in different models. For a pure charmonium state $R_{\psi\gamma}$ is predicted to be in the range of $1.2 - 15$, for a $D\overline{D}^*$ molecule in the range of $(3 - 4)\times 10^{-3}$ and for a molecule-charmonium mixture in the range of $0.5 - 5$. 
The BaBar collaboration measured the ratio $R_{\psi\gamma}$ to be $3.4\pm1.4$~\cite{BaBarXratio} while the Belle collaboration set an upper limit of $R_{\psi\gamma}<2.1$~(at 90\% C.L.)~\cite{belleXratio}. 

\begin{figure}

\centerline{\includegraphics[width=1.0\linewidth]{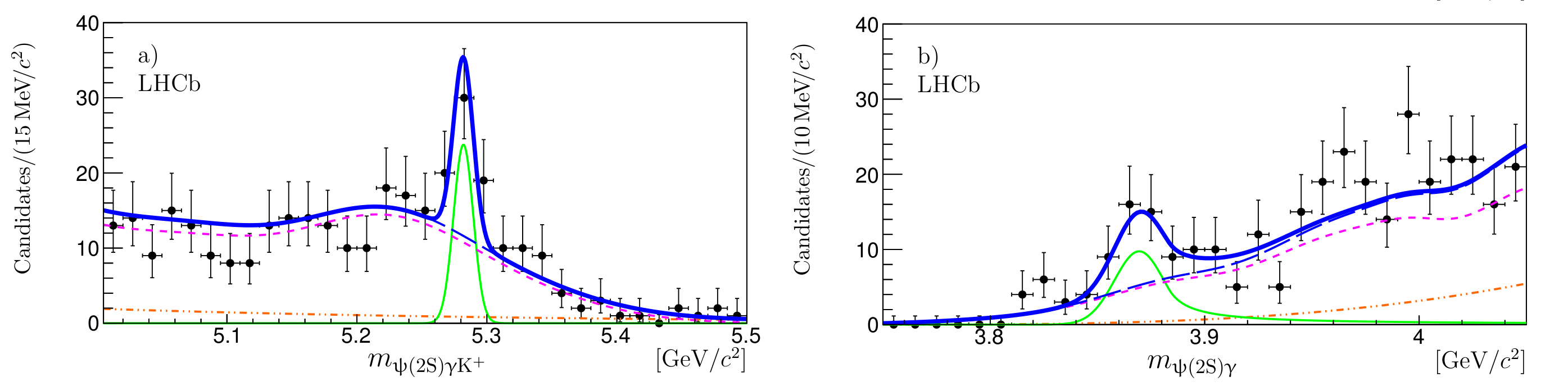}}

\caption[]{Results of the two-dimensional fit to the $B^+\rightarrow X(3872) K^+, X(3872)\rightarrow\psi(2S)\gamma$ candidates projected on a) $\psi(2S)\gamma K^+$ and b) $\psi(2S)\gamma$ invariant masses.} 
\label{fig:Xratio_comp}
\end{figure}

To resolve this discrepancy, an additional measurement has been performed by the LHCb collaboration~\cite{X3872}. The $X(3872)$ decays to $\psi\gamma$ (where $\psi$ denotes $J/\psi$ or $\psi(2S)$ meson) have been reconstructed using $B^+\rightarrow X(3872) K^+$ decays. The signal yields have been determined with two-dimensional fit to $\psi\gamma K^+$ and $\psi\gamma$ invariant masses shown in Fig.~\ref{fig:Xratio_comp}. The cross-feed from partially reconstructed $B$ meson decays have been accounted for using simulation. An evidence for the $X(3872)\rightarrow \psi(2S)\gamma$ decay has been found with significance of 4.4 standard deviations. 

The ratio $R_{\psi\gamma}$ has been measured to be $2.46 \pm 0.64 \pm 0.29$, where the first uncertainty is statistical and the second is systematic. This result is compatible with previous experiments, but more precise. This result agrees with expectations for a pure charmonium interpretation of the $X(3872)$ state and a molecular-charmonium mixture interpretations. However, it does not support a pure $D\overline{D}^*$ molecular interpretation of the $X(3872)$ state.

\section{Lifetimes of $b$ hadrons}

At leading order of Heavy Quark Expansion (HQE) theory the lifetimes of all weakly decaying $b$ hadrons are equal, with corrections appearing at $1/m_b^2$ order~\cite{HQE}. The measurements of lifetimes of the $B^+$, $B^0$ and $B_s^0$ mesons and $\Lambda_b^0$ baryon using decay modes with $J/\psi$ in a final state are presented~\cite{b-lifetimes}. The measured effective lifetimes are shown in Table~\ref{tab:lifetimes}, being the most precise single measurements up to date. The results are consistent with current world averages~\cite{PDG} and HQE predictions.

In addition, the ratio of $\Lambda_b^0$ and $B^0$ lifetimes is measured directly using ${\Lambda_b^0 \rightarrow J/\psi p K^-}$ and ${B^0 \rightarrow J/\psi \pi^+ K^-}$ decay modes~\cite{Lambdab-lifetime}. Due to kinematic similarity, the efficiencies are the same for both decays with high precision. Therefore uncertainties in acceptance efficiencies mostly cancel in the ratio. This allows for the most precise measurement of the relative $\Lambda_b$ and $B^0$ lifetimes: ${\tau_{\Lambda_b}}/{\tau_{B^0}} = 0.974 \pm 0.006 \pm 0.004$, where the first uncertainty is statistical and the second is systematic. This result is consistent with HQE predictions.

\begin{table}[t]
\caption[]{Results for the $b$ hadrons lifetimes. The first uncertainty is statistical and the second is systematic.}
\label{tab:lifetimes}
\vspace{0.4cm}
\begin{center}
\begin{tabular}{|l|c|}
\hline
decay mode & lifetime [ps] \\
\hline
$B^+\rightarrow J/\psi K^+$	& $1.637 \pm 0.004 \pm 0.003$ \\
$B^0\rightarrow J/\psi K^{*0}$	& $1.524 \pm 0.006 \pm 0.004$ \\
$B^0\rightarrow J/\psi K^0_S$	& $1.499 \pm 0.013 \pm 0.005$ \\
$\Lambda_b^0\rightarrow J/\psi \Lambda$	& $1.415 \pm 0.027 \pm 0.006$ \\
$B^0_s\rightarrow J/\psi \phi$	& $1.480 \pm 0.011 \pm 0.005$ \\
\hline

\end{tabular}
\end{center}
\end{table}

\section{$B_{c}^{+}$ physics}

The $B_{c}^{+}$ meson, composed of two heavy quarks ($\overline{b}c$), is a unique system, being the only weak decaying heavy quarkonium system. 
Prior to LHCb only the ${B_{c}^{+}\rightarrow J/\psi\mu^+\nu_{\mu}}$ and ${B_{c}^{+}\rightarrow J/\psi\pi^+}$ decay modes have been observed. LHCb has already made a great contribution to the investigation of the $B_c^+$ meson by observing many new decay modes, including the $B_c^+\rightarrow B_s^0\pi^+$ decay~\cite{Bc2Bspi}, being the first observed $B_c^+$ decay mode where $c$ quark decays. 
Previously, first observation of the $B_{c}^{+}\rightarrow J/\psi D_s^+$ decay mode allowed for the most precise measurement of the $B_c^+$ mass~\cite{Bc-mass}. 

\begin{figure}

\centerline{\includegraphics[width=1.0\linewidth]{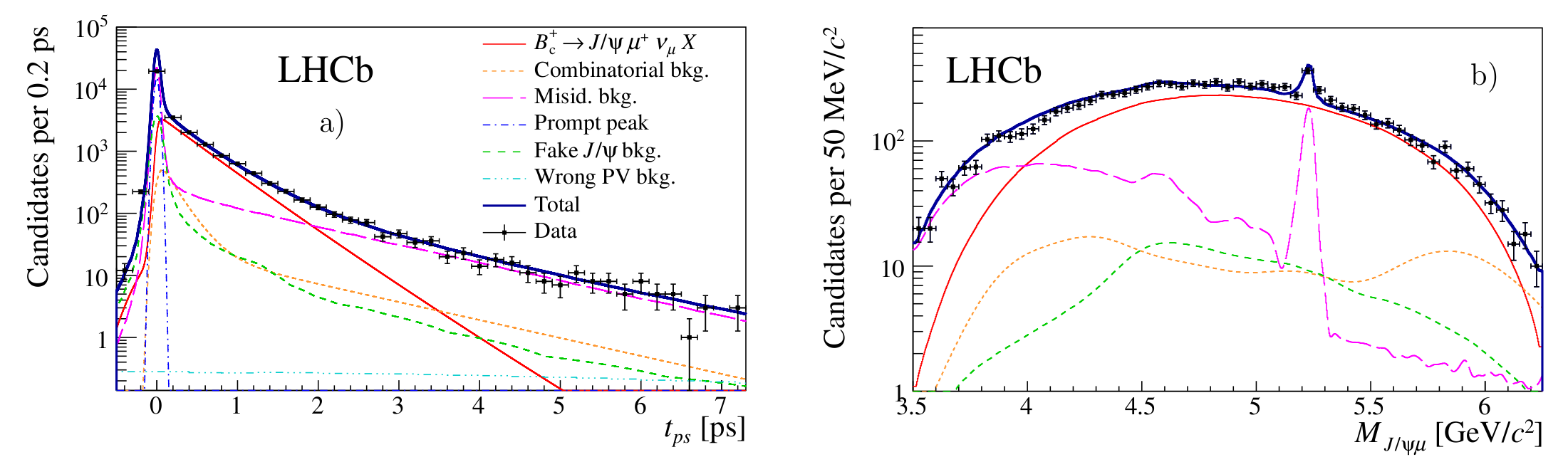}}

\caption[]{ Result of the two-dimensional fit of the $B_c^+\rightarrow J/\psi\mu^+\nu_{\mu}X$ model. Projections of the total fit function and its components are shown for (a) the pseudo-proper time and (b) the mass of the detached events ($t_{ps}>150$~fs).}
\label{fig:Bc_fit}
\end{figure}

Recently, by studying the $B_{c}^{+}\rightarrow J/\psi\mu^+\nu_{\mu}$ decays, lifetime of the $B_c^+$ has been measured~\cite{Bc-lifetime}. The lifetime has been determined from two-dimensional fit to mass of $J/\psi\mu^+$ candidates and their pseudo-proper time shown in the Fig.~\ref{fig:Bc_fit}. The background was studied using data and simulation. 
The lifetime was measured to be $\tau_{B_c^+} = 509 \pm 8 \pm 12~{\rm fs}$, where the first uncertainty is statistical and the second is systematic. The result is consistent, but more precise, than current world average~\cite{PDG}.

The decays of $B_c^+$ meson into charmonia and light hadrons are expected to be well described by the factorization approximation. 
In addition to the first observation of the ${B_c^+\rightarrow J/\psi K^+ K^- \pi^+}$ decay~\cite{Bc2jpsiKKpi}, we present the first evidence of the ${B_c^+\rightarrow J/\psi 3\pi^+2\pi^-}$ decay~\cite{Bc2jpsi5pi}. The ratios of the branching fractions to that of $B_c^+\rightarrow J/\psi \pi^+$ decays are measured to be
\bea
\frac{{\cal B}(B_c^+\rightarrow J/\psi K^+ K^- \pi^+)}{{\cal B}(B_c^+\rightarrow J/\psi \pi^+)} & = & 0.53\pm0.10\pm0.05 ~,  \\
\frac{{\cal B}(B_c^+\rightarrow J/\psi 3\pi^+2\pi^-)}{{\cal B}(B_c^+\rightarrow J/\psi \pi^+)} & = & 1.74\pm0.44\pm0.21 ~,
\eea 
\noindent where the first uncertainties are statistical and the second are systematic. The results are in agreement with theoretical predictions~\cite{Bc2jpsih_predictions} and consistent with analogous measurements in $B^0$ and $B^+$ meson decays~\cite{PDG}.

\section{Summary}

The LHCb collaboration has a rich program in spectroscopy of $b$ and $c$ hadrons, including exotic states. 
New results, based on data samples with integrated luminosities between 1.0 and 3.0~fb$^{-1}$ collected in 2011 and 2012, have been presented.
Many important and most precise results have been already achieved and more are expected in the future.

\section*{Acknowledgments}

The author wishes to thank the organizers for the financial support during the conference.

\end{document}